\documentclass[pdflatex,sn-mathphys]{sn-jnl}

\usepackage{physics}
\usepackage{amsmath,amssymb}
\DeclareMathAlphabet{\pazocal}{OMS}{zplm}{m}{n}
\DeclareRobustCommand{\stirling}{\genfrac\{\}{0pt}{}}

\jyear{2023}%

\theoremstyle{thmstyleone}%
\theoremstyle{thmstyletwo}%

\theoremstyle{thmstylethree}%

\begin{document}

\title[Evolutionary branching and consistency in human cooperation]{Evolutionary branching and consistency in human cooperation: the interplay of incentives and volunteerism in addressing collective action dilemmas}


\author*[1]{\fnm{Alina} \sur{Glaubitz}}\email{Alina.Glaubitz.GR@dartmouth.edu}

\author[1,2]{\fnm{Feng} \sur{Fu}}\email{Feng.Fu@dartmouth.edu}

\affil*[1]{\orgdiv{Department of Mathematics}, \orgname{Dartmouth College}, \city{Hanover}, \postcode{03755}, \state{NH}, \country{USA}}

\affil[2]{\orgdiv{Department of Biomedical Data Science}, \orgname{Geisel School of Medicine at Dartmouth}, \city{Lebanon}, \postcode{03756}, \state{NH}, \country{USA}}


\abstract{
Understanding the origins of volunteerism and free-riding is crucial in collective action situations where a sufficient number of cooperators is necessary to achieve shared benefits, such as in vaccination campaigns and social change movements. Despite the importance of attaining behavior consistency for successful collective action, the theoretical mechanisms behind this process remain largely elusive.
Here, we address this issue by studying the evolutionary dynamics of individual cooperativity levels in multi-round threshold public goods games. By using adaptive dynamics, we explore how individual behavior responses are shaped by incentives, with and without rewards or punishment. We demonstrate that rewarding consistent cooperators can lead to the emergence of two distinct populations: volunteers who consistently cooperate and free-riders who consistently defect. In contrast, punishing consistent defectors does not lead to similar evolutionary branching. Our results help offer insights into designing effective interventions that promote collective action and address collective risk dilemmas ranging from climate mitigation to pandemic control.

}




\maketitle

\section{Introduction}\label{sec1}

Public goods games play a crucial role in understanding the complex and dynamic nature of cooperation in human behavior~\cite{heckathorn1996dynamics}. As we face global challenges such as COVID-19, climate change, and misinformation, the importance of cooperation cannot be overstated \cite{Nowak2006,Bavel2020,Johnson2020,Brown2020,Pacheco2014,Lange2022}. From vaccination and mask wearing, to social justice movements, to health insurance, and voting, cooperation is at the forefront of each of these social dilemmas. While individuals may be tempted to act selfishly and ``free-ride'' on the contributions of others, it is only through widespread cooperation that we can achieve desired outcomes such as herd immunity, reduced carbon emissions, accessible healthcare, and fair elections that accurately represent the desires of the population. By exploring threshold public goods games~\cite{wang2009emergence}, we gain greater insight into the intricacies of human cooperation and how it interacts with incentives.

Due to its significance in a variety of real-world scenarios, exploring the behavior of individuals in public goods games has been a topic of interest for researchers (see Ref. \cite{Ledyard1995,Chaudhuri2011} for an overview). G\"achter and Fischbacher \cite{Fischbacher2001} have revealed in experiments that humans exhibit three distinct strategies when playing linear public goods games - they are either conditional cooperators, free-riders, or ``triangle'' contributors. Conditional cooperators are those who contribute based on the contributions of others, while free-riders refuse to contribute at all. Triangle contributors, on the other hand, exhibit a unique pattern of behavior. When the other players' contributions to the common pool are less than half, triangle contributors act like conditional contributors and increase their own contributions in response. However, as soon as the other players' contributions surpass half, the contributions of the triangle contributors decrease.
Gächter and Fischbacher's experiments \cite{Fischbacher2010} further provide insight into why contributions decline in iterated public goods games: Humans are imperfect conditional cooperators. They often underestimate the contributions made by their fellow players, causing them to fall short in their own efforts to contribute to the common good. Yet, the question remains: what drives individuals to adopt these different strategies? This mystery calls for further investigation, and the use of innovative approaches like the adaptive dynamics framework and differing neighborhood sizes and contributions have shown promising results in revealing the evolutionary branching of a population into cooperators and defectors. The study of non-linear public goods games, particularly threshold public goods games, is crucial in understanding the dynamic changes in human cooperation and incentives.

Therefore, Doebeli et al. \cite{Doebeli2004} took a closer look at nonlinear payoff functions in a snow-drift game. They used the adaptive dynamics framework to showcase how such functions could split a population into cooperators and defectors. Similarly, Johnson et al. \cite{Johnson2021} revealed how differing the size of neighborhoods and contributions can lead to the division of cells into cooperative and non-cooperative groups. These models both have a common characteristic, in that the payoff does not linearly depend on the contributions. In fact, many real-world situations, such as threshold public goods games, can be best described by non-linear models. This is why it is crucial to delve deeper into how dynamics change in these types of games.

Looking for ways to increase cooperation levels in a population has been a widely-discussed topic of research, with different mechanisms such as volunteering \cite{Hauert2002,Hauert2007} or insurance \cite{Zhang2015} being suggested. However, one mechanism that has proven to be highly effective is incentives. The evolution of decentralized rewards and punishment has been studied in different contexts, with both rewards and punishment shown to increase contributions in public goods games~\cite{szolnoki2013effectiveness,chen2015first}. But among these, punishment is considered a stronger motivator \cite{Fischbacher2001,Henrich2001,Bolton1995}. The impact of rewarding cooperators and punishing defectors in public goods games has been analyzed by Sigmund et al. \cite{Sigmund2001} using bifurcation analysis, including and excluding reputation effects. Punishment caused a social equilibrium to arise, but rewarding led to highly unpredictable dynamics. Meanwhile, Hilbe et al. \cite{Hilbe2010} evaluated the evolution of opportunism in a donation game, revealing how rewarding can evolve in a population of defectors, while punishment can sustain cooperation in a population that is already cooperative. Dreber et al. \cite{Dreber2008} conducted experiments and found that ``Winners don't punish.'' While social punishment alone can sustain cooperation, it cannot be sustained in the presence of anti-social punishment \cite{Rand2011}. Reputation effects, however, can establish cooperation by promoting punishing and rewarding \cite{Santos2011,Hilbe2012,Pal2022}. In conclusion, the combination of incentives and reputation effects can play a vital role in establishing and maintaining cooperation in a population.

In addition, third-party punishment \cite{Andreoni2003} as well as coordinated incentives on preventing free riding has been studied \cite{Chen2015,Garcia2019,Sigmund2010}. The role of centralized incentives in promoting cooperation has also been examined through both experiments and models \cite{Baldassarri2011,Micheli2021,Dong2016}. Similar to decentralized incentives, centralized incentives can encourage cooperation in a population, with punishment having a greater impact than rewards \cite{Diekmann2015}. Furthermore, the presence of an endogenous choice between centralized rewards and punishment greatly enhances the level of cooperation \cite{Sutter2010,Dannenberg2020}.

In this work, we examine the impact of various incentives on cooperation in a multi-round threshold public goods game. When rewarding cooperative players, our findings reveal the emergence of two distinct groups of players - ``volunteers'' who consistently cooperate and ``free-riders'' who do not. This evolutionary branching is facilitated when rewards are convex in the number of rounds of cooperation 
as it incentivizes consistency. Furthermore, we observe that while punishing consistent defectors can initiate volunteer behavior in a population, rewarding consistent volunteers leads to widespread adoption of cooperative behavior once a certain threshold is reached. Finally, our examination of the model in a broader strategy space reveals that while the branching of volunteers and free-riders remains present, it becomes unsustainable and oscillates between periods of branching and convergence back to a monomorphic population state.

\section{Model and Methods}\label{sec2}

\subsection{Base Model without Incentives}\label{sec3}

\begin{figure}
    \centering
    \includegraphics[width=\textwidth]{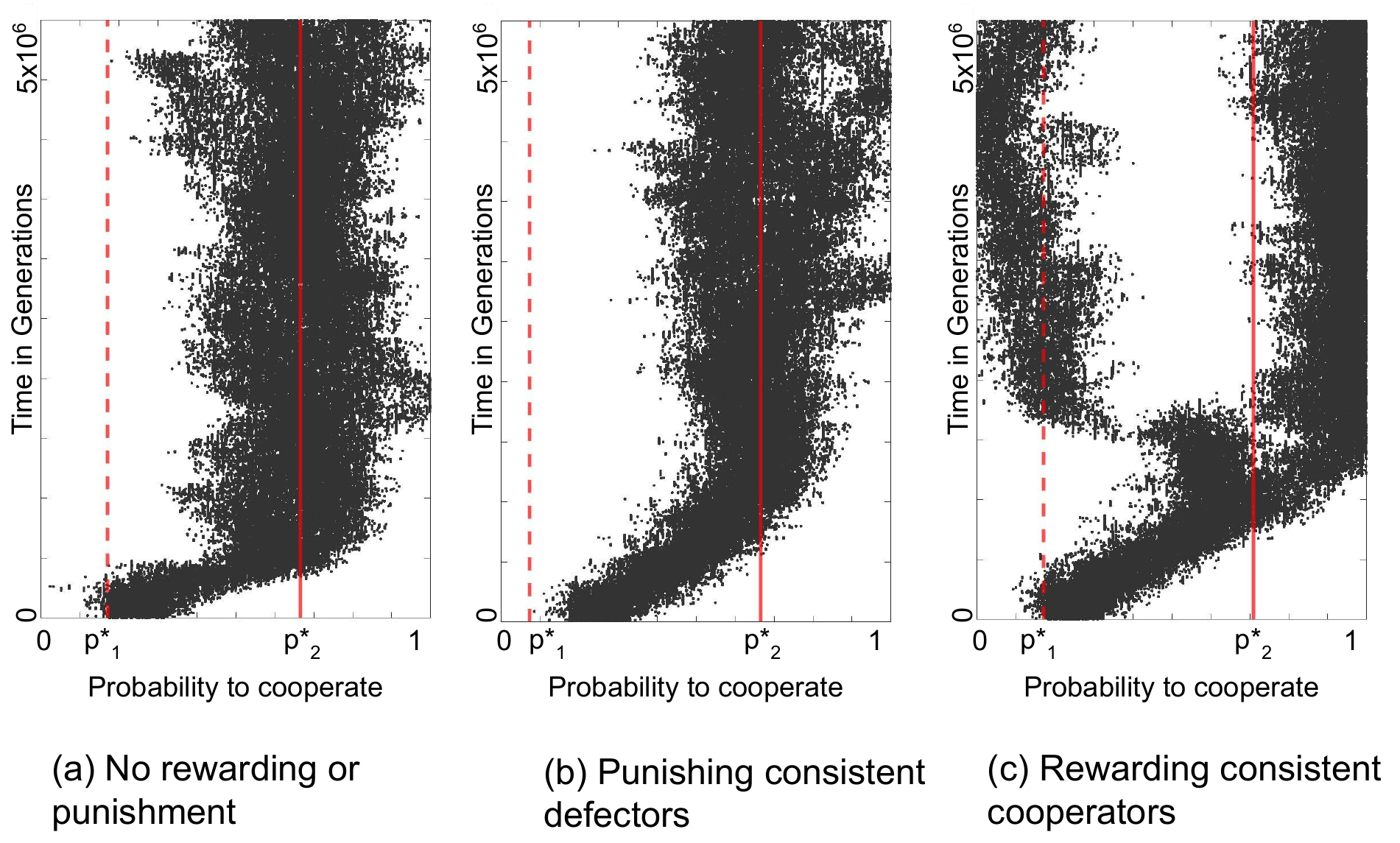}
    \caption{Evolutionary branching of volunteers and free-riders is possible under rewarding. Simulations of the population dynamics (a) without either rewards or punisment, (b) with rewarding consistent cooperators, and (c) with punishing consistent defectors. Parameters are population size $M = 800$, group size $N = 6, \mathcal{M}_C = 3, R= 6, C = 1, \lambda = 5, \mu = 0.01, \sigma = 0.075, T_{\text{end}} = 5\times10^6$. The vertical solid red lines are convergent stable strategies, and the vertical dashed red lines are repeller strategies.}
    \label{fig:simulations}
\end{figure}

In this study, we analyze a multi-round game where $N$ players play together for $K$ rounds, each making a choice to cooperate or defect in each round. The outcome of the game is determined by the cooperation of enough players, resulting in a common benefit $B>0$ for all players if the threshold $\mathcal{M}_C$ is reached. Individual rewards $R$ or punishments $P$ are assigned based on the player's cooperative behavior over $K$ rounds. Our research objective is to understand the effect of different incentive mechanisms on cooperation dynamics. When consistent cooperation is rewarded, we observe a evolutionary branching into volunteer players who always cooperate and free-riders who never do. This branching is absent without rewards or when consistent defection is punished (see Figure \ref{fig:simulations}).

To analyze this model, we utilize the adaptive dynamics framework \cite{Geritz1997}, which is based on the assumption of a large population that begins as monomorphic, meaning all individuals exhibit the same trait. This population evolves through the strategy space through small mutations and social imitation, with soft selection determined by the payoff of the public goods game. To understand this evolution, we consider the fitness of a rare mutant with strategy $p_m$ in a population of resident players with strategy $p_r$. The fitness is given by a linear equation in $p_m$ and a polynomial of degree $N-1$ in $p_r$, namely
\begin{align}
    \text{E}[p_m,p_r] &= 
    B\sum_{k=\mathcal{M}_C}^{N-1} {N-1 \choose k} p_r^k(1-p_r)^{N-1-k} \\
    &+ p_m (B {N-1 \choose \mathcal{M}_C-1} p_r^{\mathcal{M}_C-1}(1-p_r)^{N-\mathcal{M}_C}-C).
\end{align}

The concept of the selection gradient is central to understanding the evolution of the cooperation probability $p$. The selection gradient, denoted as $D(p)$, is calculated as the derivative of the expected value of $p_m$ and $p_r$ with respect to $p_m$, evaluated at $p_m = p_r = p$,
 \begin{align}
 D(p) &= \left. \pdv{E[p_m,p_r]}{p_m}\right\vert_{p_m = p_r=p} \\
 &= B {N-1 \choose \mathcal{M}_C-1} p^{\mathcal{M}_C-1}(1-p)^{N-\mathcal{M}_C} - C
 \end{align}
It is noteworthy that the selection gradient determines the evolution of $p$ since its rate of change is given by $p' = D(p)$. When $D(p)$ is positive, $p$ increases over time, and when $D(p)$ is negative, $p$ decreases over time.
The equilibria of the evolutionary dynamics are identified as the roots of $D(p)$ and referred to as singular strategies. In this model, the selection gradient is represented as a polynomial of degree $N-1$.

\begin{figure}
    \centering
    \includegraphics[width=0.9\textwidth]{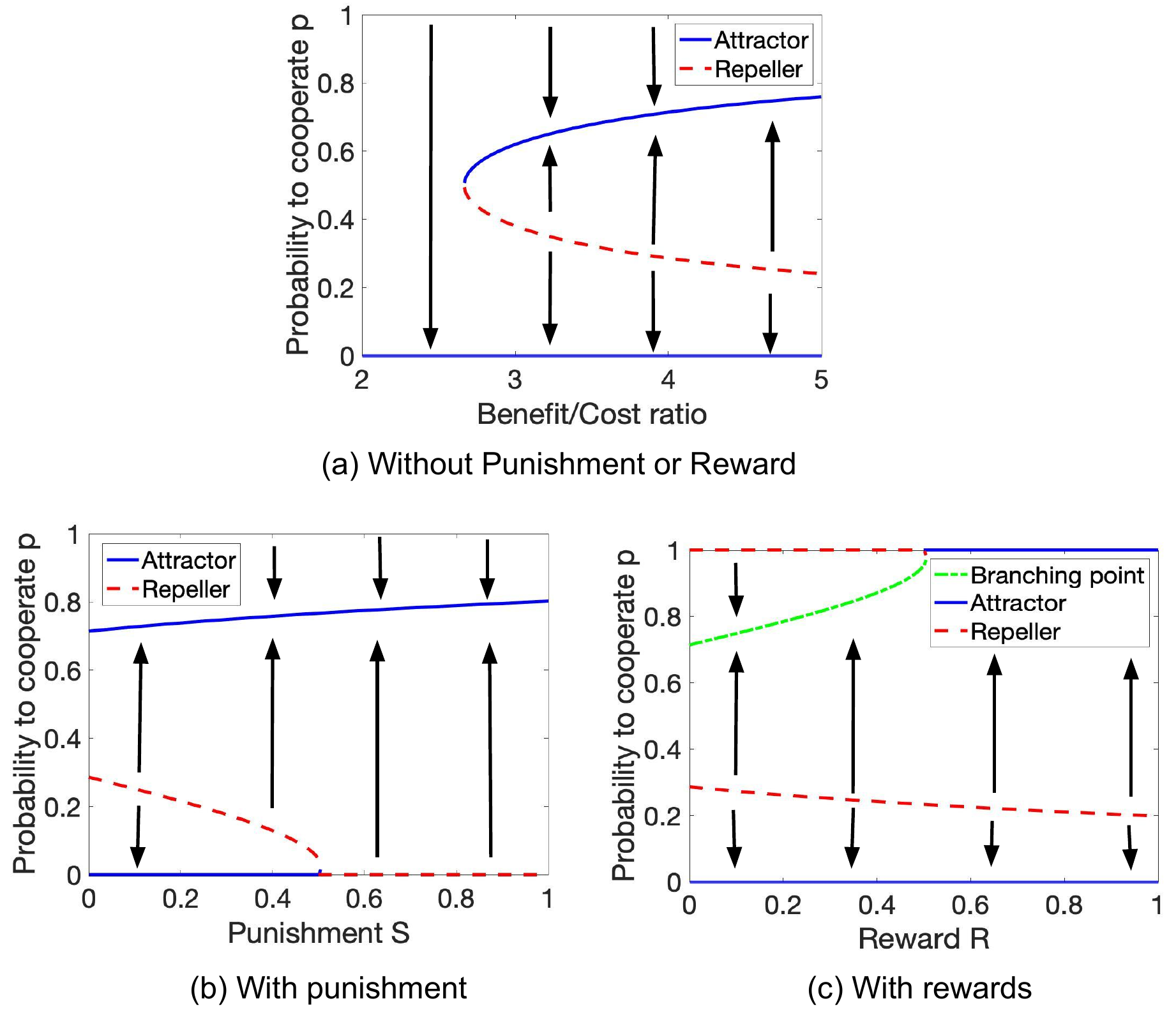}
    \caption{Bifurcation diagrams under different incentive mechanisms. (a) Equilibria in the public goods game (without rewards or punishment) $N = 5, \mathcal{M}_C = 3, R = 2, C = 1$. (b) Equilibria for the model with punishment for any cooperation with
    $N = 8, \mathcal{M}_C = 7, R = 20, C = 1,\lambda = 5$ and different punishment $S$. (c) Equilibria for the model with rewards for consistent cooperators with
    $N = 8, \mathcal{M}_C = 2, R = 12, C = 1,\lambda = 5$ and different rewards $R$.}
    \label{fig:singular}
\end{figure}

The polynomial $D$ can be analyzed by examining three cases:
\begin{itemize}
    \item[(1)] $\mathcal{M}_C = 1$: In this case, the maximum of the polynomial $D$ is at $p=0$ and the value at $p=0$ is positive. This means that there is one stable root $0 < p_2^* < 1$ that attracts the population's strategy towards it. When the probability of the other players cooperating is low, it's best to cooperate, but when the other players are likely to cooperate, it's best to defect. The population will evolve towards the social equilibrium $p_2^*$ from any initial strategy.
    \item[(2)] $\mathcal{M}_C \in \{2,\dots,N-1\}$: In this case, $0<\frac{\mathcal{M}_C-1}{N-1} <1$ and if the benefit to cost ratio is large enough there exist two roots $0<p_1^*<p_2^*<1$, one convergent stable equilibrium $p_2^*$ and one repelling equilibrium $p_1^*$. Here, it is the best strategy to cooperate when the other players are cooperating with a probability between $p_1^*$ and $p_2^*$. For an illustration of the singular strategies in case (2), see Figure \ref{fig:singular} (a). Here we see that for small benefit to cost ratios ($B/C$), only the anti-social equilibrium exists. As $B/C$ increases, the two equilibrium strategies $0<p_1^*<p_2^*<1$ evolve. Then, for any inital probability to cooperate between 0 and $p_1^*$, we evolve toward the anti-social equilibrium. For any initial probability between $p_1^*$ and 1, we evolve to the social equilibrium $p_2^*$. 
    \item[(3)] $\mathcal{M}_C = N$: In this case, the maximum of the polynomial $D$ is at $p = 1$ and there is only one root $0 < p_1^* < 1$. This root acts as a repeller, meaning that if the initial probability to cooperate is low, the population will evolve towards the anti-social equilibrium, but if the initial probability is high, the population will evolve towards full cooperation. It makes sense to cooperate only when the probability of all other players cooperating is high, but not when it is low.

\end{itemize}

Once the the social equilibrium is reached in cases (1) and (2), the further evolution is determined by the invasion fitness, $E[p_m,p_2^*]$, at the social equilibrium $p_2^*$. In this scenario, both cooperation and defection yield the same result in a monomorphic population of $p_2^*$ players, as the invasion fitness is constant. This means that any mutant strategy will receive the same payoff as the resident strategy. However, if the average of cooperation in the population increases, defection will have a higher payoff, causing the population to shift back to adopting $p_2^*$. Conversely, if the population average decreases below $p_2^*$, cooperation will become more advantageous, leading to an increase in the average again. As a result, once the population reaches the social equilibrium, it will oscillate around this average (as depicted in Figure \ref{fig:simulations} (a)).

\section{Results}

\subsection{Evolutionary Branching: Comparing Rewards versus Punishment} \label{ss:ex}

Let us now incorporate into the base model two distinct reward and punishment mechanisms that are based on behavioral consistency. Players participate in a minimum of $K\geq2$ rounds and receive a reward based on the number of rounds in which they exhibit cooperative behavior. The expected reward $R$ for a given strategy $p$ is calculated as follows:
$\text{E}[\text{R}] = \sum_{j=0}^K {K \choose j} p_m^k (1-p_m)^{K-j} a_j$, where $a_j,j \in{1,\dots,K},$ represents the reward received for cooperating in $j$ rounds.
For the purpose of this analysis, we shall consider a special case where players receive a reward only if they cooperate in all rounds of play, commonly referred to as an ``all or nothing'' scenario (see SI for the results on the coevolution of rewarding size). The reward function in this case is defined as follows:
\begin{equation}
    a_j = \begin{cases} a, & j = K, \\ 0, & \text{otherwise.}\end{cases}
\end{equation}
The expected reward gained by an individual player is expressed as $\text{E}[\text{R}]=ap_m^K$. From this, the selection gradient is calculated as $D(p) = B {N-1 \choose \mathcal{M}_C-1} p^{\mathcal{M}_C-1}(1-p)^{N-\mathcal{M}_C} - C + a K p_m^{K-1}$. In comparison to the model without incentives, the equilibrium $p_2^*$ in this model exhibits a distinct nature, as it becomes a fitness minimum with a second derivative of $\left. \pdv[2]{E[p_m,p_2^*]}{p_m} \right\vert_{p_m = p_2^*} = aK(K-1)p_m^{K-2}>0$. As a result, once the population evolves to $p_2^*$, it splits into two subpopulations, with one subpopulation consistently cooperating ($p=1$) and the other never cooperating ($p=0$)  (as depicted in Figure \ref{fig:simulations}(a)). This process is known as "evolutionary branching," and we refer to the cooperative subpopulation as \textbf{volunteers} and the non-cooperative subpopulation as \textbf{free-riders}. The equilibrium $p_2^*$ is referred to as a branching point.

The polynomial equation $D(p)=0$ may have zero, one, two, or three roots $0<p^*_1<p^*_2<p^*_3<1$, if $\mathcal{M}_C\neq 1,N$. These roots represent different equilibria (see SI for further details), and the corresponding bifurcation diagrams can be observed in Figure \ref{fig:singular} (b).
It is worth noting that the rewards for consistent cooperation have a significant impact on the behavior of $p_2^*$ compared to $p_1^*$. As the reward $R$ increases, a saddle-node bifurcation, combined with a transcritical bifurcation, occurs. For small rewards $a<C/K$, $p_2^*$ is a branching point. On the other hand, for large rewards $a>C/K$, full cooperation ($p=1$) becomes a convergently stable state, and as soon as the cooperation probability exceeds $p_1^*$, the population evolves towards full cooperation.

Additionally, when $\mathcal{M}_C=1$, the unique equilibrium $p_2^*$ acts as a branching point and we observe a similar bifurcation behavior as in the case $\mathcal{M}_C\neq 1,N$. In contrast, when $\mathcal{M}_C=N$, the behavior observed is similar to that in the scenario where rewards are not present. The behavior of the population is not significantly influenced by the presence of rewards.

The introduction of punishment for consistent free-riding modifies the game as follows: consistent defectors' payoffs are reduced by an amount equal to the punishment $P$. The expected punishment is given by $\text{E}[\text{P}] = - S(1-p_m)^K$. The selection gradient in this case is given by $D(p)= B {N-1 \choose \mathcal{M}_C-1} p^{\mathcal{M}_C-1}(1-p)^{N-\mathcal{M}_C} - C + SK(1-p_m)^{K-1}$. The main difference in this model from the others is that $p_2^*$ becomes a fitness maximum, meaning it is evolutionarily stable. This leads to the population evolving to the social equilibrium, following the same pattern as in the original model without rewards or punishment.

This is due to the fact that $\left. \pdv[2]{E[p_m,p_2^*]}{p_m} \right\vert_{p_m = p_2^*} = -SK(K-1)p_m^{K-2}<0$.

For $\mathcal{M}_C$ values between 1 and $N$, and for small values of $S<C/K$, the model has one repeller $p_1^*$ and one attractor $p_2^*$. However, as $S$ increases above $C/K$, the repeller exhibits a transcritical bifurcation with the anti-social equilibrium $p=0$, which becomes a repeller. As a result, the population evolves towards the social equilibrium $p_2^*$ for any initial probability of cooperation. A bifurcation diagram of these dynamics can be found in Figure \ref{fig:singular} (c).

The results of our simulations for the different reward and punishment mechanisms explored in Subsection \ref{ss:ex} can be visualized in Figure \ref{fig:simulations}. The findings highlight that: (a) In the absence of rewards or punishment, the dynamics of the game converge towards the CSS, where defection and cooperation are equally viable strategies in terms of payoffs. (b) The imposition of punishment for consistent defection results in the CSS becoming a fitness maximum, promoting evolutionary stability. (c) The introduction of rewards for cooperative behavior leads to a bifurcation of the population into volunteers and free-riders as the CSS becomes a fitness minimum (see SI for the pairwise invasibility plots).

\subsection{Conditional strategies: Moving from unconditional strategies}

Our observations indicate that the long-term behavior of the model is contingent upon the strategy space under consideration. In larger strategy spaces, we observe instances of evolutionary branching. However, as the dimensionality of the strategy space increases to three or more, the branching becomes unsustainable and is characterized by recurrent cycles of branching and collapsing.

For a more detailed analysis (see Section 5 of Extended Methods), we fix the number of rounds $K=2$ and reward only those players who consistently cooperate in both rounds. Each player's strategy is characterized by three probabilities: $(p_0, p_C, p_D)$. $p_0$ is the initial probability to cooperate, $p_C$ is the probability to cooperate after having cooperated in the first round, and $p_D$ is the probability to cooperate after having defected in the first round.
Our analysis reveals that the strategy $(p_0,p_C,p_D) = (p_2^*, 1, 0)$ is convergence stable, but not evolutionarily stable. Specifically, this strategy is a fitness minimum in the direction of $p_0 = p_C$, a fitness maximum in the direction of $p_0 = -p_C$, and neutral in the direction of $p_D$. As a result, we observe evolutionary branching of $p_0$ into cooperators and defectors, with the population splitting in the direction of $p_0=p_C$. However, after branching into volunteers and free-riders, one of the branches collapses, leading the population to evolve back to $(p_0,p_C,p_D) = (p_2^*, 1, 0)$, where it branches again. The speed at which the branches collapse and develop is influenced by the mutation size $\sigma$ and frequency $\mu$.

When a correlation exists between cooperation in the first and second round, the behavior of unconditional strategies is maintained, resulting in the evolution of $p$ towards $p_{(3,cor)} =(p_2^*,p_2^*,p_2^*)$. This evolution results in the branching of the population into two distinct groups: volunteers ($p_{(3,1)}=(1,1,1)$) and free-riders ($p_{(3,2)} = (0,0,0)$). Our findings indicate that even when the correlation is not perfect, simple unconditional strategies can preserve the branching in volunteers and free-riders, even in larger strategy spaces.

\begin{figure}
    \centering
    \includegraphics[width=0.7\textwidth]{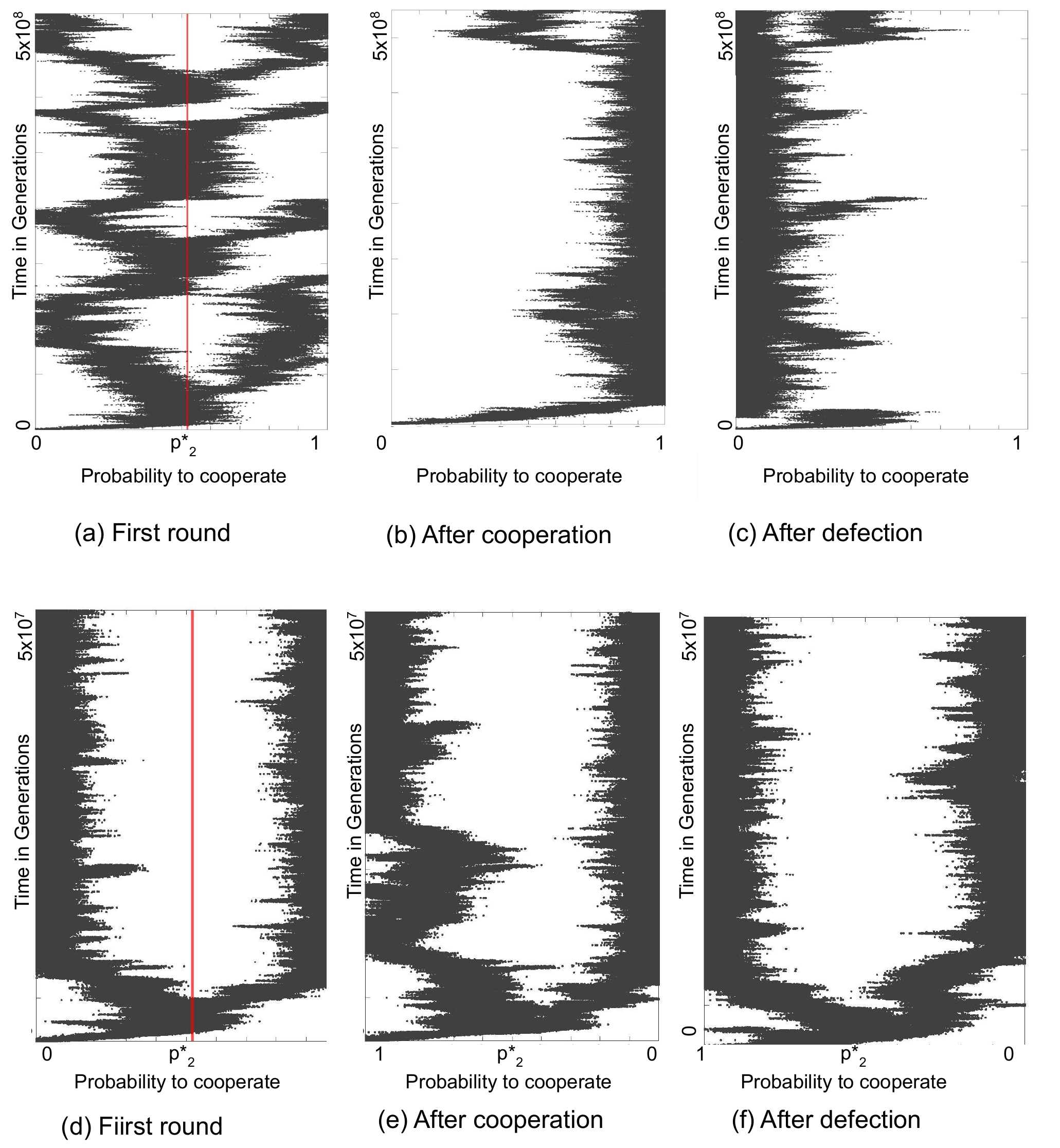}
    \caption{Intermittent cycles of branching and collapsing. (a),(b),(c) For a larger strategy space in two rounds, we still see evolutionary branching. Now, however, the branches collapse again before branching out a second time, and so on. (d),(e),(f) show the same when there is a correlation $\Sigma$ between the strategies. Parameters are population size $M = 1500$, group size $N = 3, \mathcal{M}_C = 1, R= 3, C = 1, \lambda = 2, \mu = 0.01, \sigma = 0.02$, (a) - (c) $T_{\text{end}} = 5\times10^8$, and (d) - (f) $T_{\text{end}} = 5\times 10^7$. The vertical solid red lines are convergence stable strategies but not evolutionarily stable, and other boundary equilibria are not shown for clarity.}
    \label{fig:larger}
\end{figure}

This nontrivial phenomenon is illustrated in Figure \ref{fig:larger}. For further exploration, the adaptive dynamics of this model are discussed in detail in the Extended Methods section.

\subsection{Conditions of General Rewarding Structure}

\begin{figure}
    \centering
    \includegraphics[width=0.7\textwidth]{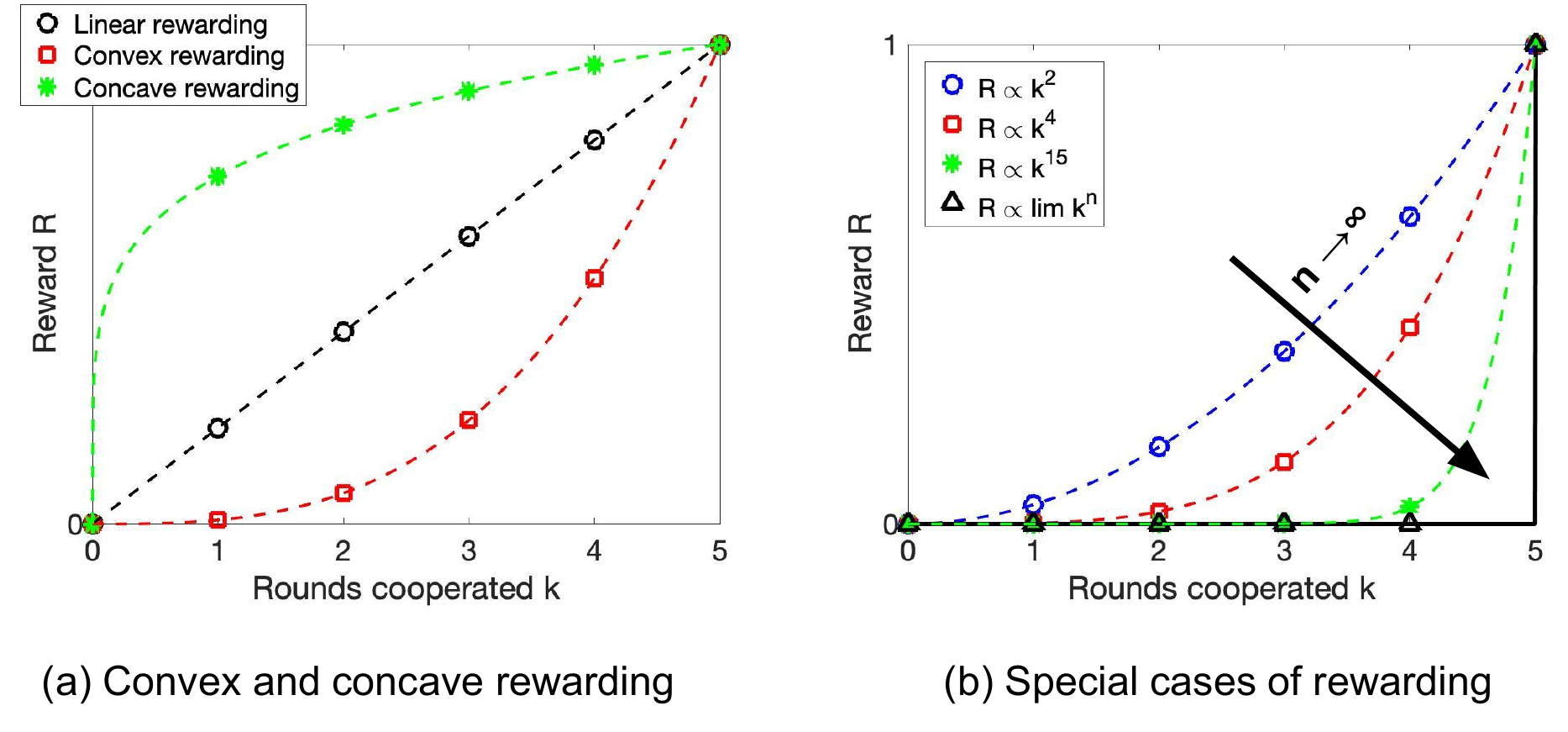}
    \caption{Impact of rewarding structure. (a) Comparison of convex and concave rewarding mechanisms. (b) Convex rewarding that causes evolutionary branching in volunteers and free-riders, including the limit case.}
    \label{fig:convex}
\end{figure}

We highlight that the rewarding mechanism discussed in Section \ref{ss:ex} is not the the only one to cause evolutionary branching. In order to observe evolutionary branching, a singular strategy needs to satisfy two conditions:
\begin{itemize}
    \item[(i)]
    $\dv{D(p)}{p}<0$, i.\,e.\ convergence stable,
    \item[(ii)]
    $\dv[2]{\text{E}[p_m,p_r]}{p_m}>0$, i.\,e.\ a fitness minimum and not evolutionarily stable.
\end{itemize}
Now, in this model, these are equivalent to
\begin{align} \label{eq:IR_cond}
        0&< \left.\dv[2]{\text{E}[\text{R}]}{p_m}\right\vert_{p_m=p_r=p_2^*} \\&< B {N-1 \choose \mathcal{M}_C-1} {p^*_2}^{\mathcal{M}_C-2}(1-{p^*_2})^{N-\mathcal{M}_C-1}(\mathcal{M}_C-1-(N-1)p^*_2).
\end{align}
We note here that for $a_i = 0, \forall i = 0,\dots, N,$ the model becomes the model without rewarding. Moreover, the expectation of $R$ continuously depends on $a_i, i \in \{0,\dots,N\}$. Hence as long as the right hand side of the above inequality is satisfied, there exists $a>0$ such that $a$R satisfies \eqref{eq:IR_cond}. If players play $K=2$ rounds, the condition becomes 
$\text{E}[\text{R}] = (1-p_m)^{2} a_0 + 2p_m(1-p_m) a_1 + p_m^2 a_2$
and therefore $\dv[2]{p_m}\text{E}[\text{R}] = 2a_0-4a_1+2a_2.$
Hence, in order for this to cause evolutionary branching, $a_0+a_2>2a_1$ needs to be satisfied, which means that consistency needs to be rewarded compared to inconsistency in order to observe branching.
In general, the condition can be interpreted as the rewards being convex. We can see an illustration of these different conditions of rewarding structure in Figure \ref{fig:convex} (a). 

Here, we consider the special case $a_n = a\left(\frac{n}{K}\right)^c, n = 0,\dots,K, c \in \mathbb{N}$. In this scenario, the expected value of the reward is given as the $c$-th moment of the binomial distribution, in particular, $\text{E}[\text{R}] = \text{E}_{p_m}[aX^c] = \sum_{j=0}^c \stirling{c}{j} K^{\underline{j}} p_m^j,$
where $\stirling{c}{j}$ is the Stirling number of the second kind \cite{Graham1988}, and $K^{\underline{j}} = K(K-1)\cdots (K-j+1)$ is the falling factorial. Note here, that both the Stirling number and the falling factorial are positive (for $j>0$). Therefore, for $c\geq 2$, \eqref{eq:IR_cond} is satisfied. Discussed in Section \ref{ss:ex} is the limit case of this model as $c \to \infty$, also see Figure \ref{fig:convex} (b). 

\section{Discussions and Conclusion}
\label{sec12}

In this work, we explore the impact of rewards and punishments on the evolution of cooperation in a threshold public goods game. Our results demonstrate that rewarding can trigger evolutionary branching in both volunteers and free-riders, whereas punishment is ineffective in doing so. This finding is novel and differs from previous research which has investigated the effect of both decentralized \cite{Sigmund2001,Hilbe2010,Andreoni2003}, coordinated \cite{Chen2015,Garcia2019} and centralized incentives \cite{Diekmann2015,Sutter2010,Dannenberg2020} in a variety of games, including evolutionary branching in a continuous snowdrift game with a nonlinear payoff function \cite{Doebeli2004} and a nonlinear public goods game with varying neighborhood size \cite{Johnson2021}. Hence, our model is an illustration of how negative feedback stabilizes the behavior of the model, while positive feedback is self-amplifying and can lead to bistable systems.

Behavioral consistency is an important aspect of human behavior~\cite{sih2004behavioral}, which has been empirically observed in behavioral experiments~\cite{wedekind1996human,molleman2014consistent,haesevoets2018behavioural} and also explored in evolutionary modeling studies~\cite{wolf2011coevolution,mcnamara2020behavioural}. Our results show that behavioral consistency can be the consequence of evolutionary branching due to the presence of certain rewarding incentive structure as aforementioned. This is an imporant insight arising from the present study that can inform future incentive design for fostering consistent cooperation in repeated interactions.    

In conclusion, our work provides insight into the evolutionary origins of volunteerism and free-riding in multi-round threshold public goods games. Using adaptive dynamics, we reveal how individual optimum behavior responses are influenced by incentives, with and without rewards or punishment. Our findings demonstrate that rewarding consistent cooperators can lead to the emergence of two distinct populations: volunteers who consistently cooperate and free-riders who consistently defect. However, punishing consistent defectors, while promoting cooperation, does not result in evolutionary branching. These results improve our understanding of the interplay between incentive structures and behavior consistency, shedding new light onto promoting consistent cooperation in collective action scenarios.

\section{Extended Methods}\label{sec11}

In this section, we delve into the dynamics of a larger strategy space. Our analysis focuses on a two-round game, where players have the ability to adjust their strategy in the second round based on their behavior in the first round. We characterize each strategy by the probabilities $(p_0,p_C,p_D)$, with $p_0$ representing the probability of cooperation in the first round, $p_C$ representing the probability of cooperation in the second round after a cooperation in the first round, and $p_D$ representing the probability of cooperation in the second round after a defection in the first round.

Surprisingly, the introduction of this added complexity still leads to branching dynamics, as the branches merge and re-emerge repeatedly. This cycle begins with $p_C$ approaching 1 and $p_D$ approaching 0, while $p_0$ converges towards $p_0^*$. This sets the stage for further branching, which is then followed by a collapse, and so on. These intriguing dynamics are a subject of ongoing research and hold promise for a deeper understanding of the underlying mechanisms driving cooperative behavior in multi-round interactions.

In a model with individual rewarding, the payoff of this extended model is given by
\begin{align*}
    &E\left[ \begin{pmatrix} p_{m,0} \\ p_{m,C} \\ p_{m,D} \end{pmatrix}, \begin{pmatrix} p_{r,0} \\ p_{r,C} \\ p_{r,D} \end{pmatrix} \right] =
    -p_{m,0} C + B\sum_{k=\mathcal{M}_C}^{N-1} {N-1 \choose k} p_{r,0}^k(1-p_{r,0})^{N-1-k} \\
    &+ p_{m,0} B {N-1 \choose \mathcal{M}_C-1} p_{r,0}^{\mathcal{M}_C-1}(1-p_{r,0})^{N-\mathcal{M}_C} \\
    &+ B\sum_{k=\mathcal{M}_C}^{N-1} {N-1 \choose k} p_{r,0}^k(1-p_{r,0})^{N-1-k} \sum_{j = \mathcal{M}_C}^{N-1} \\
    &\sum_{m = 0}^j {k \choose m} {N-1-k \choose j-m} p_{r,C}^m (1-p_{r,C})^{k-m} p_{r,D}^{j-m} (1-p_{r,D})^{N-1-k-(j-m)} \\
    &+ (p_{m,0}p_{m,C} + (1-p_{m,0})p_{m,D})\left(B\sum_{k=\mathcal{M}_C}^{N-1} {N-1 \choose k} p_{r,0}^k(1-p_{r,0})^{N-1-k} \right.\\
    &\left.\sum_{m = 0}^{\mathcal{M}_C-1} {k \choose m} {N-1-k \choose \mathcal{M}_C-1-m} p_{r,C}^m (1-p_{r,C})^{k-m} p_{r,D}^{\mathcal{M}_C-1-m} (1-p_{r,D})^{N-k-(\mathcal{M}_C-m)} -C \right) \\
    &+ a_2 p_{m,0}p_{m,C} + a_1 (p_{m,0}(1-p_{m,C})+(1-p_{m,0})p_{m,D}).
\end{align*}
In particular, the second derivative of the invasion fitness with respect to the mutant strategies is given by
\[
    \pdv[2]{E[p_m,p_r]}{p_m} = \begin{pmatrix} 0 & a_2 - a_1 & - a_1 \\   a_2 - a_1 & 0 & 0 \\ -a_1 & 0 & 0 \end{pmatrix}
\]
Specifically, this matrix is indefinite with eigenvalues (i) 0, (ii) $\pm \sqrt{a_2^2-2a_1a_2 + a_1^2}$. The according eigenvectors are given by
\[
    \text{(i)} \begin{pmatrix} 0 \\ a_1 \\ a_2-a_1 \end{pmatrix} \text{respectively  (ii)} \begin{pmatrix} \pm (a_2-a_1) \\ a_1-a_2 \\ a_1 \end{pmatrix}
\]

In this analysis, it is indicated that the singular points of the system can be classified as saddle points. This is because they are the best response to themselves in the direction of $A_1 = \begin{pmatrix} a_1-a_2, & a_1-a_2, & a_1 \end{pmatrix}^T$, but in the direction of $A_2 = \begin{pmatrix} a_2-a_1, & a_1-a_2, & a_1 \end{pmatrix}^T$, any other response is a better response. Consequently, mutations that occur in the direction of $A_1$ will lead to a collapse of the branches, whereas mutations that occur in the direction of $A_2$ will result in branching. The presence of correlation between the traits $p_0, p_C$ and $p_D$ can prevent the collapse of the branches, with the extreme case of perfect correlation being discussed previously. When $a_1$ is set to zero and $a_2$ is fixed at a value greater than zero, the eigenvalues take the form of (i) zero and (ii) $\pm 1$ with corresponding eigenvectors (i) $\begin{pmatrix} 0 \ 0 \ 1 \end{pmatrix}$ and (ii) $\begin{pmatrix} \pm 1 \ -1 \ 0 \end{pmatrix}$.

\backmatter

\bmhead{Supplementary information}
The paper has an online Supplementary Information that details further analyses of the model. 



\bmhead{Acknowledgments}
F.F. gratefully acknowledges support from the Bill \& Melinda Gates Foundation (award no. OPP1217336), the NIH COBRE Program (grant no.1P20GM130454), and the Neukom CompX Faculty Grant. We thank Christian Hilbe and Moshe Hoffmann for helpful discussions.


\section*{Declarations}


\begin{itemize}
\item Funding\\
This work is supported by the Bill \& Melinda Gates Foundation (award no. OPP1217336).

\item Conflict of interest/Competing interests\\
The authors declare no conflicting interests.

\item Ethics approval \\
Not applicable.

\item Consent to participate\\
Not applicable.

\item Consent for publication\\
All authors reviewed and approved the manuscript for final publication. 
\item Availability of data and materials\\
All data and materals have been included in the main text and the Supplementary Information.

\item Code availability \\
Source code to reproduce the results in the paper is available upon reasonable request. 

\item Authors' contributions\\
A.G. \& F.F. conceived the project, performed the reseach, and wrote the manuscript. A.G. performed simulations and plotted all figures. 
\end{itemize}







\begin{appendices}




\end{appendices}


\bibliography{sn-bibliography}


\begin{thebibliography}{46}
\ifx \bisbn   \undefined \def \bisbn  #1{ISBN #1}\fi
\ifx \binits  \undefined \def \binits#1{#1}\fi
\ifx \bauthor  \undefined \def \bauthor#1{#1}\fi
\ifx \batitle  \undefined \def \batitle#1{#1}\fi
\ifx \bjtitle  \undefined \def \bjtitle#1{#1}\fi
\ifx \bvolume  \undefined \def \bvolume#1{\textbf{#1}}\fi
\ifx \byear  \undefined \def \byear#1{#1}\fi
\ifx \bissue  \undefined \def \bissue#1{#1}\fi
\ifx \bfpage  \undefined \def \bfpage#1{#1}\fi
\ifx \blpage  \undefined \def \blpage #1{#1}\fi
\ifx \burl  \undefined \def \burl#1{\textsf{#1}}\fi
\ifx \doiurl  \undefined \def \doiurl#1{\url{https://doi.org/#1}}\fi
\ifx \betal  \undefined \def \betal{\textit{et al.}}\fi
\ifx \binstitute  \undefined \def \binstitute#1{#1}\fi
\ifx \binstitutionaled  \undefined \def \binstitutionaled#1{#1}\fi
\ifx \bctitle  \undefined \def \bctitle#1{#1}\fi
\ifx \beditor  \undefined \def \beditor#1{#1}\fi
\ifx \bpublisher  \undefined \def \bpublisher#1{#1}\fi
\ifx \bbtitle  \undefined \def \bbtitle#1{#1}\fi
\ifx \bedition  \undefined \def \bedition#1{#1}\fi
\ifx \bseriesno  \undefined \def \bseriesno#1{#1}\fi
\ifx \blocation  \undefined \def \blocation#1{#1}\fi
\ifx \bsertitle  \undefined \def \bsertitle#1{#1}\fi
\ifx \bsnm \undefined \def \bsnm#1{#1}\fi
\ifx \bsuffix \undefined \def \bsuffix#1{#1}\fi
\ifx \bparticle \undefined \def \bparticle#1{#1}\fi
\ifx \barticle \undefined \def \barticle#1{#1}\fi
\bibcommenthead
\ifx \bconfdate \undefined \def \bconfdate #1{#1}\fi
\ifx \botherref \undefined \def \botherref #1{#1}\fi
\ifx \url \undefined \def \url#1{\textsf{#1}}\fi
\ifx \bchapter \undefined \def \bchapter#1{#1}\fi
\ifx \bbook \undefined \def \bbook#1{#1}\fi
\ifx \bcomment \undefined \def \bcomment#1{#1}\fi
\ifx \oauthor \undefined \def \oauthor#1{#1}\fi
\ifx \citeauthoryear \undefined \def \citeauthoryear#1{#1}\fi
\ifx \endbibitem  \undefined \def \endbibitem {}\fi
\ifx \bconflocation  \undefined \def \bconflocation#1{#1}\fi
\ifx \arxivurl  \undefined \def \arxivurl#1{\textsf{#1}}\fi
\csname PreBibitemsHook\endcsname

\bibitem{heckathorn1996dynamics}
\begin{botherref}
\oauthor{\bsnm{{Heckathorn, Douglas D}}}:
The dynamics and dilemmas of collective action.
American sociological review,
250--277
(1996).
\doiurl{10.2307/2096334}
\end{botherref}
\endbibitem

\bibitem{Nowak2006}
\begin{barticle}
\bauthor{\bsnm{{Nowak, Martin}}}:
\batitle{Five rules for the evolution of cooperation}.
\bjtitle{Science}
\bvolume{314}(\bissue{5805}),
\bfpage{1560}--\blpage{1563}
(\byear{2006}).
\doiurl{10.1126/science.1133755}
\end{barticle}
\endbibitem

\bibitem{Bavel2020}
\begin{barticle}
\bauthor{\bsnm{{Bavel, J.J.V.}}},
\bauthor{\bsnm{{Baicker, K.}}},
\bauthor{\bsnm{{Boggio, P.S.}}},
\bauthor{\bparticle{et} \bsnm{al.}}:
\batitle{Using social and behavioural science to support covid-19 pandemic
  response}.
\bjtitle{Nat Hum Behav}
\bvolume{4},
\bfpage{460}--\blpage{471}
(\byear{2020}).
\doiurl{10.1038/s41562-020-0884-z}
\end{barticle}
\endbibitem

\bibitem{Johnson2020}
\begin{botherref}
\oauthor{\bsnm{{Johnson, Tim}}},
\oauthor{\bsnm{{Dawes, Christopher}}},
\oauthor{\bsnm{{Fowler, James}}},
\oauthor{\bsnm{{Smirnov, Oleg}}}:
Slowing covid-19 transmission as a social dilemma: Lessons for government
  officials from interdisciplinary research on cooperation.
Journal of Behavioral Public Administration
\textbf{3}(1)
(2020).
\doiurl{10.30636/jbpa.31.150}
\end{botherref}
\endbibitem

\bibitem{Brown2020}
\begin{barticle}
\bauthor{\bsnm{{Brown, Gordon}}},
\bauthor{\bsnm{{Susskind, Daniel}}}:
\batitle{International cooperation during the covid-19 pandemic}.
\bjtitle{Oxford Review of Economic Policy}
\bvolume{36}(\bissue{1}),
\bfpage{64}--\blpage{76}
(\byear{2020}).
\doiurl{10.1093/oxrep/graa025}
\end{barticle}
\endbibitem

\bibitem{Pacheco2014}
\begin{barticle}
\bauthor{\bsnm{{Pacheco, Jorge M.}}},
\bauthor{\bsnm{{Vasconcelos, Vítor V. }}},
\bauthor{\bsnm{{Santos, Francisco C.}}}:
\batitle{Climate change governance, cooperation and self-organization}.
\bjtitle{Physics of Life Reviews}
\bvolume{11}(\bissue{4}),
\bfpage{573}--\blpage{586}
(\byear{2014}).
\doiurl{10.1016/j.plrev.2014.02.003}
\end{barticle}
\endbibitem

\bibitem{Lange2022}
\begin{barticle}
\bauthor{\bsnm{{Van Lange, Paul A.M.}}},
\bauthor{\bsnm{{Rand, David G.}}}:
\batitle{Human cooperation and the crises of climate change, covid-19, and
  misinformation}.
\bjtitle{Annual Review of Psychology}
\bvolume{73}(\bissue{1}),
\bfpage{379}--\blpage{402}
(\byear{2022}).
\doiurl{10.1146/annurev-psych-020821-110044}
\end{barticle}
\endbibitem

\bibitem{wang2009emergence}
\begin{barticle}
\bauthor{\bsnm{{Wang, Jing}}},
\bauthor{\bsnm{{Fu, Feng}}},
\bauthor{\bsnm{{Wu, Te}}},
\bauthor{\bsnm{{Wang, Long}}}:
\batitle{Emergence of social cooperation in threshold public goods games with
  collective risk}.
\bjtitle{Physical Review E}
\bvolume{80}(\bissue{1}),
\bfpage{016101}
(\byear{2009}).
\doiurl{10.1103/PhysRevE.80.016101}
\end{barticle}
\endbibitem

\bibitem{Ledyard1995}
\begin{bchapter}
\bauthor{\bsnm{{{Ledyard, John O.}}}}:
\bctitle{Public goods: A survey of experimental research}.
In: \beditor{\bsnm{{Roth, Alvin E.}}},
\beditor{\bsnm{{Kagel, John H.}}} (eds.)
\bbtitle{The Handbook of Experimental Economics},
pp. \bfpage{111}--\blpage{194}.
\bpublisher{Princeton University Press},
\blocation{Princeton}
(\byear{1995}).
\bcomment{Chap. 2}
\end{bchapter}
\endbibitem

\bibitem{Chaudhuri2011}
\begin{barticle}
\bauthor{\bsnm{{{Chaudhuri, Ananish}}}}:
\batitle{Sustaining cooperation in laboratory public goods experiments: A
  selective survey of the literature}.
\bjtitle{Experimental Economics}
\bvolume{14}(\bissue{1}),
\bfpage{47}--\blpage{83}
(\byear{2014}).
\doiurl{10.1007/s10683-010-9257-1}
\end{barticle}
\endbibitem

\bibitem{Fischbacher2001}
\begin{barticle}
\bauthor{\bsnm{Fischbacher}, \binits{U.}},
\bauthor{\bsnm{Gächter}, \binits{S.}},
\bauthor{\bsnm{Fehr}, \binits{E.}}:
\batitle{Are people conditionally cooperative? evidence from a public goods
  experiment}.
\bjtitle{Economics Letters}
\bvolume{71}(\bissue{3}),
\bfpage{397}--\blpage{404}
(\byear{2001}).
\doiurl{10.1016/S0165-1765(01)00394-9}
\end{barticle}
\endbibitem

\bibitem{Fischbacher2010}
\begin{barticle}
\bauthor{\bsnm{{Fischbacher, Urs}}},
\bauthor{\bsnm{{Gächter, Simon}}}:
\batitle{Social preferences, beliefs, and the dynamics of free riding in public
  goods experiments}.
\bjtitle{American Economic Review}
\bvolume{100}(\bissue{1}),
\bfpage{541}--\blpage{56}
(\byear{2010}).
\doiurl{10.1257/aer.100.1.541}
\end{barticle}
\endbibitem

\bibitem{Doebeli2004}
\begin{barticle}
\bauthor{\bsnm{{Doebeli, Michael}}},
\bauthor{\bsnm{{Hauert, Christoph}}},
\bauthor{\bsnm{{Killingback, Timothy}}}:
\batitle{The evolutionary origin of cooperators and defectors}.
\bjtitle{Science}
\bvolume{306}(\bissue{5697}),
\bfpage{859}--\blpage{862}
(\byear{2004}).
\doiurl{10.1126/science.1101456}
\end{barticle}
\endbibitem

\bibitem{Johnson2021}
\begin{barticle}
\bauthor{\bsnm{{Johnson, Brian}}},
\bauthor{\bsnm{{Altrock, Philipp M.}}},
\bauthor{\bsnm{{Kimmel, Gregory J.}}}:
\batitle{Two-dimensional adaptive dynamics of evolutionary public goods games:
  finite-size effects on fixation probability and branching time}.
\bjtitle{R. Soc. open sci.}
\bvolume{8},
\bfpage{8210182}
(\byear{2021}).
\doiurl{10.1098/rsos.210182}
\end{barticle}
\endbibitem

\bibitem{Hauert2002}
\begin{barticle}
\bauthor{\bsnm{{ Hauert, Christoph}}},
\bauthor{\bsnm{{ De Monte, Silvia}}},
\bauthor{\bsnm{{ Hofbauer, Josef}}},
\bauthor{\bsnm{{ Sigmund, Karl}}}:
\batitle{Volunteering as red queen mechanism for cooperation in public goods
  games}.
\bjtitle{Science}
\bvolume{296}(\bissue{5570}),
\bfpage{1129}--\blpage{1132}
(\byear{2002}).
\doiurl{10.1126/science.1070582}
\end{barticle}
\endbibitem

\bibitem{Hauert2007}
\begin{barticle}
\bauthor{\bsnm{{Hauert, Christoph}}},
\bauthor{\bsnm{{Traulsen, Arne}}},
\bauthor{\bsnm{{Brandt, Hannelore}}},
\bauthor{\bsnm{{ Nowak, Martin}}},
\bauthor{\bsnm{{Sigmund, Karl }}}:
\batitle{Via freedom to coercion: The emergence of costly punishment}.
\bjtitle{Science}
\bvolume{316}(\bissue{5833}),
\bfpage{1905}--\blpage{1907}
(\byear{2007}).
\doiurl{10.1126/science.1141588}
\end{barticle}
\endbibitem

\bibitem{Zhang2015}
\begin{barticle}
\bauthor{\bsnm{{Zhang, Jianlei}}},
\bauthor{\bsnm{{Zhang, Chunyan}}},
\bauthor{\bsnm{{Cao, Ming}}}:
\batitle{How insurance affects altruistic provision in threshold public goods
  games}.
\bjtitle{Sci Rep}
\bvolume{5},
\bfpage{9098}
(\byear{2015}).
\doiurl{10.1038/srep09098}
\end{barticle}
\endbibitem

\bibitem{szolnoki2013effectiveness}
\begin{barticle}
\bauthor{\bsnm{{Szolnoki, Attila}}},
\bauthor{\bsnm{{Perc, Matja{\v{z}}}}}:
\batitle{Effectiveness of conditional punishment for the evolution of public
  cooperation}.
\bjtitle{Journal of theoretical biology}
\bvolume{325},
\bfpage{34}--\blpage{41}
(\byear{2013}).
\doiurl{10.1016/j.jtbi.2013.02.008}
\end{barticle}
\endbibitem

\bibitem{chen2015first}
\begin{barticle}
\bauthor{\bsnm{{Chen, Xiaojie}}},
\bauthor{\bsnm{{Sasaki, Tatsuya}}},
\bauthor{\bsnm{{Br{\"a}nnstr{\"o}m, {\AA}ke}}},
\bauthor{\bsnm{{Dieckmann, Ulf}}}:
\batitle{First carrot, then stick: how the adaptive hybridization of incentives
  promotes cooperation}.
\bjtitle{Journal of the royal society interface}
\bvolume{12}(\bissue{102}),
\bfpage{20140935}
(\byear{2015}).
\doiurl{10.1098/rsif.2014.0935}
\end{barticle}
\endbibitem

\bibitem{Henrich2001}
\begin{barticle}
\bauthor{\bsnm{{Henrich, Joseph}}},
\bauthor{\bparticle{et} \bsnm{al.}}:
\batitle{In search of homo economicus: Behavioral experiments in 15 small-scale
  societies}.
\bjtitle{The American economic review}
\bvolume{91}(\bissue{2}),
\bfpage{73}--\blpage{78}
(\byear{2001}).
\doiurl{10.1257/aer.91.2.73}
\end{barticle}
\endbibitem

\bibitem{Bolton1995}
\begin{barticle}
\bauthor{\bsnm{{Bolton, Gary E.}}},
\bauthor{\bsnm{{Zwick, Rami}}}:
\batitle{Anonymity versus punishment in ultimatum bargaining}.
\bjtitle{Games and Economic Behavior}
\bvolume{10}(\bissue{1}),
\bfpage{95}--\blpage{121}
(\byear{1995}).
\doiurl{10.1006/game.1995.1026}
\end{barticle}
\endbibitem

\bibitem{Sigmund2001}
\begin{barticle}
\bauthor{\bsnm{{Sigmund, Karl}}},
\bauthor{\bsnm{{Hauert, Christoph}}},
\bauthor{\bsnm{{Nowak, Martin}}}:
\batitle{Reward and punishment}.
\bjtitle{PNAS}
\bvolume{98}(\bissue{19}),
\bfpage{10757}--\blpage{62}
(\byear{2001}).
\doiurl{10.1073/pnas.161155698}
\end{barticle}
\endbibitem

\bibitem{Hilbe2010}
\begin{botherref}
\oauthor{\bsnm{{Hilbe, Christian}}},
\oauthor{\bsnm{{Sigmund, Karl}}}:
Incentives and opportunism: from the carrot to the stick.
Proc. R. Soc. B,
2772427--2433
(2010).
\doiurl{10.1098/rspb.2010.0065}
\end{botherref}
\endbibitem

\bibitem{Dreber2008}
\begin{barticle}
\bauthor{\bsnm{{Dreber, Anna}}},
\bauthor{\bsnm{{Rand, David}}},
\bauthor{\bsnm{{Fudenberg, Drew}}},
\bauthor{\bsnm{{Nowak, Martin}}}:
\batitle{Winners don’t punish}.
\bjtitle{Nature}
\bvolume{452},
\bfpage{348}--\blpage{351}
(\byear{2008}).
\doiurl{10.1038/nature06723}
\end{barticle}
\endbibitem

\bibitem{Rand2011}
\begin{botherref}
\oauthor{\bsnm{{Rand, David}}},
\oauthor{\bsnm{{Nowak, Martin}}}:
The evolution of antisocial punishment in optional public goods games.
Nat Commun
\textbf{2}(434)
(2011).
\doiurl{10.1038/ncomms1442}
\end{botherref}
\endbibitem

\bibitem{Santos2011}
\begin{botherref}
\oauthor{\bsnm{{Santos, Miguel dos}}},
\oauthor{\bsnm{{Rankin, Daniel J.}}},
\oauthor{\bsnm{{Wedekind, Claus}}}:
The evolution of punishment through reputation.
Proc R Soc B,
278371--377
(2011).
\doiurl{10.1098/rspb.2010.1275}
\end{botherref}
\endbibitem

\bibitem{Hilbe2012}
\begin{botherref}
\oauthor{\bsnm{{Hilbe, Christian}}},
\oauthor{\bsnm{{Traulsen, Arne}}}:
Emergence of responsible sanctions without second order free riders, antisocial
  punishment or spite.
Sci Rep
\textbf{2}(458)
(2012).
\doiurl{10.1038/srep00458}
\end{botherref}
\endbibitem

\bibitem{Pal2022}
\begin{barticle}
\bauthor{\bsnm{{Pal, Saptarshi}}},
\bauthor{\bsnm{{Hilbe, Christian}}}:
\batitle{Reputation effects drive the joint evolution of cooperation and social
  rewarding}.
\bjtitle{Nat Commun}
\bvolume{13},
\bfpage{5928}
(\byear{2022}).
\doiurl{10.1038/s41467-022-33551-y}
\end{barticle}
\endbibitem

\bibitem{Andreoni2003}
\begin{barticle}
\bauthor{\bsnm{{Andreoni, James}}},
\bauthor{\bsnm{{Harbaugh, William}}},
\bauthor{\bsnm{{Vesterlund, Lise}}}:
\batitle{The carrot or the stick: Rewards, punishments, and cooperation}.
\bjtitle{American Economic Review}
\bvolume{93}(\bissue{3}),
\bfpage{893}--\blpage{902}
(\byear{2003}).
\doiurl{10.1257/000282803322157142}
\end{barticle}
\endbibitem

\bibitem{Chen2015}
\begin{barticle}
\bauthor{\bsnm{{Chen, Xiaojie}}},
\bauthor{\bsnm{{ Szolnoki, Attila}}},
\bauthor{\bsnm{{ Perc, Matjaz}}}:
\batitle{Competition and cooperation among different punishing strategies in
  the spatial public goods game}.
\bjtitle{Phys. Rev. E}
\bvolume{92},
\bfpage{012819}
(\byear{2015}).
\doiurl{10.1103/PhysRevE.92.012819}
\end{barticle}
\endbibitem

\bibitem{Garcia2019}
\begin{botherref}
\oauthor{\bsnm{{García, Julián}}},
\oauthor{\bsnm{{Traulsen, Arne}}}:
Evolution of coordinated punishment to enforce cooperation from an unbiased
  strategy space.
J. R. Soc. Interface,
2019012720190127
(2019).
\doiurl{10.1098/rsif.2019.0127}
\end{botherref}
\endbibitem

\bibitem{Sigmund2010}
\begin{barticle}
\bauthor{\bsnm{{Sigmund, Karl}}},
\bauthor{\bsnm{{De Silva, Hannelore}}},
\bauthor{\bsnm{{Traulsen, Arne}}},
\bauthor{\bsnm{{Hauert, Christoph}}}:
\batitle{Social learning promotes institutions for governing the commons}.
\bjtitle{Nature}
\bvolume{466}(\bissue{7308}),
\bfpage{861}--\blpage{3}
(\byear{2010}).
\doiurl{10.1038/nature09203}
\end{barticle}
\endbibitem

\bibitem{Baldassarri2011}
\begin{barticle}
\bauthor{\bsnm{{Baldassarri, Delia}}},
\bauthor{\bsnm{{Grossman, Guy}}}:
\batitle{Centralized sanctioning and legitimate authority promote cooperation
  in humans}.
\bjtitle{PNAS}
\bvolume{108}(\bissue{27}),
\bfpage{11023}--\blpage{11027}
(\byear{2011}).
\doiurl{10.1073/pnas.1105456108}
\end{barticle}
\endbibitem

\bibitem{Micheli2021}
\begin{botherref}
\oauthor{\bsnm{{Micheli, Leticia}}},
\oauthor{\bsnm{{Stallen, Mirre}}},
\oauthor{\bsnm{{Sanfey, Alan G.}}}:
The effect of centralized financial and social incentives on cooperative
  behavior and its underlying neural mechanisms.
Brain Sci
\textbf{11}(3)
(2021).
\doiurl{10.3390/brainsci11030317}
\end{botherref}
\endbibitem

\bibitem{Dong2016}
\begin{barticle}
\bauthor{\bsnm{{Dong, Yali}}},
\bauthor{\bsnm{{Zhang, Boyu}}},
\bauthor{\bsnm{{Tao, Yi}}}:
\batitle{The dynamics of human behavior in the public goods game with
  institutional incentives}.
\bjtitle{Sci Rep}
\bvolume{6},
\bfpage{28809}
(\byear{2016}).
\doiurl{10.1038/srep28809}
\end{barticle}
\endbibitem

\bibitem{Diekmann2015}
\begin{barticle}
\bauthor{\bsnm{{Diekmann, A.}}},
\bauthor{\bsnm{{Przepiorka, W.}}}:
\batitle{Punitive preferences, monetary incentives and tacit coordination in
  the punishment of defectors promote cooperation in humans}.
\bjtitle{Sci Rep}
\bvolume{5},
\bfpage{10321}
(\byear{2015}).
\doiurl{10.1038/srep10321}
\end{barticle}
\endbibitem

\bibitem{Sutter2010}
\begin{barticle}
\bauthor{\bsnm{{Sutter, Matthias}}},
\bauthor{\bsnm{{Haigner, Stefan}}},
\bauthor{\bsnm{{Kocher, Martin}}}:
\batitle{Choosing the carrot or the stick? endogenous institutional choice in
  social dilemma situations}.
\bjtitle{The Review of Economic Studies}
\bvolume{77}(\bissue{4}),
\bfpage{1540}--\blpage{1566}
(\byear{2010}).
\doiurl{10.1111/j.1467-937X.2010.00608.x}
\end{barticle}
\endbibitem

\bibitem{Dannenberg2020}
\begin{barticle}
\bauthor{\bsnm{{Dannenberg, Alexander}}},
\bauthor{\bsnm{{Gallier, Carlo}}}:
\batitle{The choice of institutions to solve cooperation problems: a survey of
  experimental research}.
\bjtitle{Experimental Economics}
\bvolume{23},
\bfpage{716}--\blpage{749}
(\byear{2020}).
\doiurl{10.1007/s10683-019-09629-8}
\end{barticle}
\endbibitem

\bibitem{Geritz1997}
\begin{barticle}
\bauthor{\bsnm{{Geritz, Stefan A. H.}}},
\bauthor{\bsnm{{Metz, J. A. J.}}},
\bauthor{\bsnm{{Kisdi, \'Eva}}},
\bauthor{\bsnm{{Mesz\'ena, G\'eza}}}:
\batitle{Dynamics of adaptation and evolutionary branching}.
\bjtitle{Phys. Rev. Lett.}
\bvolume{78}(\bissue{10}),
\bfpage{2024}--\blpage{2027}
(\byear{1997}).
\doiurl{10.1103/PhysRevLett.78.2024}
\end{barticle}
\endbibitem

\bibitem{Graham1988}
\begin{botherref}
\oauthor{\bsnm{{Graham, Ronald L.}}},
\oauthor{\bsnm{{ Knuth, Donald E.}}},
\oauthor{\bsnm{{Patashnik, Oren}}}:
Concrete Mathematics,
p. 244
\end{botherref}
\endbibitem

\bibitem{sih2004behavioral}
\begin{barticle}
\bauthor{\bsnm{{Sih, Andrew}}},
\bauthor{\bsnm{{Bell, Alison}}},
\bauthor{\bsnm{{Johnson, J Chadwick}}}:
\batitle{Behavioral syndromes: an ecological and evolutionary overview}.
\bjtitle{Trends in ecology \& evolution}
\bvolume{19}(\bissue{7}),
\bfpage{372}--\blpage{378}
(\byear{2004}).
\doiurl{10.1016/j.tree.2004.04.009}
\end{barticle}
\endbibitem

\bibitem{wedekind1996human}
\begin{barticle}
\bauthor{\bsnm{{Wedekind, Claus}}},
\bauthor{\bsnm{{Milinski, Manfred}}}:
\batitle{Human cooperation in the simultaneous and the alternating prisoner's
  dilemma: Pavlov versus generous tit-for-tat.}
\bjtitle{Proceedings of the National Academy of Sciences}
\bvolume{93}(\bissue{7}),
\bfpage{2686}--\blpage{2689}
(\byear{1996}).
\doiurl{10.1073/pnas.93.7.2686}
\end{barticle}
\endbibitem

\bibitem{molleman2014consistent}
\begin{barticle}
\bauthor{\bsnm{{Molleman, Lucas}}},
\bauthor{\bsnm{{Van den Berg, Pieter}}},
\bauthor{\bsnm{{Weissing, Franz J}}}:
\batitle{Consistent individual differences in human social learning
  strategies}.
\bjtitle{Nature Communications}
\bvolume{5}(\bissue{1}),
\bfpage{3570}
(\byear{2014}).
\doiurl{10.1038/ncomms4570}
\end{barticle}
\endbibitem

\bibitem{haesevoets2018behavioural}
\begin{barticle}
\bauthor{\bsnm{{Haesevoets, Tessa}}},
\bauthor{\bsnm{{Reinders, Folmer, Chris}}},
\bauthor{\bsnm{{Bostyn, Dries H}}},
\bauthor{\bsnm{{Van Hiel, Alain}}}:
\batitle{Behavioural consistency within the prisoner's dilemma game: the role
  of personality and situation}.
\bjtitle{European Journal of Personality}
\bvolume{32}(\bissue{4}),
\bfpage{405}--\blpage{426}
(\byear{2018}).
\doiurl{10.1002/per.2158}
\end{barticle}
\endbibitem

\bibitem{wolf2011coevolution}
\begin{barticle}
\bauthor{\bsnm{{Wolf, Max}}},
\bauthor{\bsnm{{Van Doorn, G Sander}}},
\bauthor{\bsnm{{Weissing, Franz J}}}:
\batitle{On the coevolution of social responsiveness and behavioural
  consistency}.
\bjtitle{Proceedings of the Royal Society B: Biological Sciences}
\bvolume{278}(\bissue{1704}),
\bfpage{440}--\blpage{448}
(\byear{2011}).
\doiurl{10.1098/rspb.2010.1051}
\end{barticle}
\endbibitem

\bibitem{mcnamara2020behavioural}
\begin{barticle}
\bauthor{\bsnm{{McNamara, John M}}},
\bauthor{\bsnm{{Barta, Zoltan}}}:
\batitle{Behavioural flexibility and reputation formation}.
\bjtitle{Proceedings of the Royal Society B}
\bvolume{287}(\bissue{1939}),
\bfpage{20201758}
(\byear{2020}).
\doiurl{10.1098/rspb.2020.1758}
\end{barticle}
\endbibitem

\end{thebibliography}


\newpage
\begin{center}
    {\Large{{Supplementary Material for\\
``Evolutionary branching and consistency in human cooperation: the interplay of incentives and volunteerism in addressing collective action dilemmas''}}}
\medskip

    {Alina Glaubitz and Feng Fu}

\end{center}

\section{Coevolution of Rewarding or Punishment}

\begin{figure}
    \centering
    \includegraphics[width=.98\textwidth]{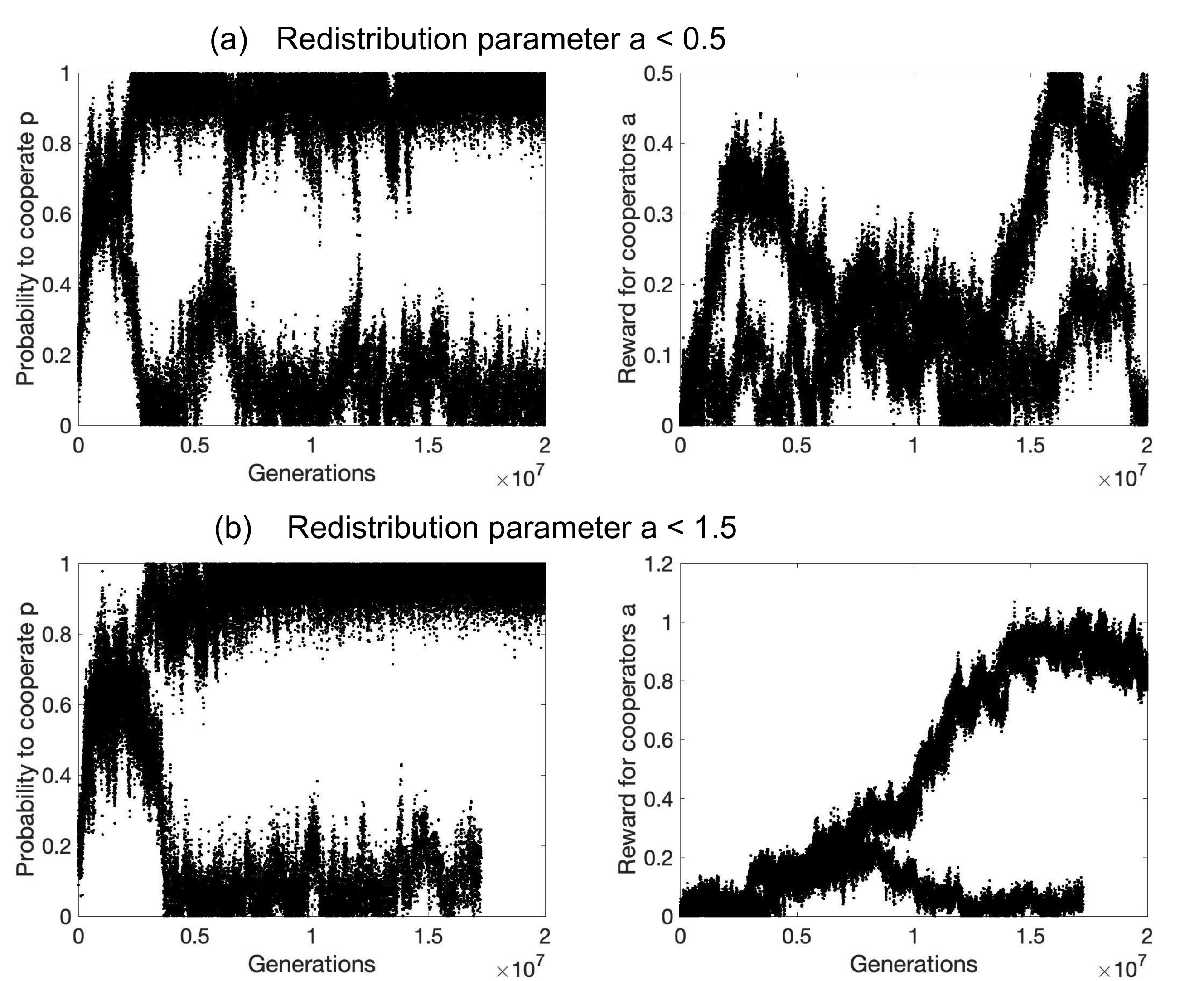}
    \caption{Simulations of the game with rewarding evolving simultaneously. Parameters are $M = 1000, N = 5, \mathcal{M}_C = 2, B = 6, C = 1,K = 4,P = 0,\mu = 0.01, \sigma_1 = 0.075,\sigma_2 = 0.03,T_{\text{end}} = 20000000$.}
    \label{fig:rewardingEvolution}
\end{figure}

We look at an extension of the game that includes the evolution of rewarding and punishing. In particular, players can decide if the benefit obtained by the game gets redistributed to consistent cooperators respectively if free-riders get a smaller proportion of the benefit. So, in this scenario, the reward and punishment are taken from the reward that the population obtained as a whole through enough cooperation. Similar to before, rewarding causes evolutionary branching while punishment does not. However, if there are no bounds on the amount redistributed, the population evolves to all volunteers over time. An illustration can be seen in Figure \ref{fig:rewardingEvolution}. Note that this is not a fitness maximum. 
Through random drift, rewarding evolves in the population, as in a monomorphic population, all players have the same expected payoff. However, as soon as the population branches into volunteers and free-riders, the redistribution parameter for volunteers increases, while the redistribution for free-riders goes to $R=0$. So, the primary branching into volunteers and free-riders causes a secondary evolutionary branching in redistributors and non-redistributors. 
Finally, if $R$ can evolve without bound from above, i.\,e.\ $R$ becomes larger than the cost of cooperation, the population evolves to full cooperation and the free-rider branch dies out, until random drift decreases $R$ to a value small enough for defectors being able to invade again.
If $R$ is bounded from above this behavior does not occur.

In this extension of the game, the expected payoff of a mutant player with strategy $(p_m,R_m)$ (volunteering with probability $p_m$ and redistributing $R_m$ of the reward to the consistent cooperators) in a population of resident players with strategy $(p_r,R_r)$ is given by
\begin{align*}
    &E[(p_m,R_m),(p_r,R_r)] 
    = -p_m C \\
    &+ \left(R-\frac{1}{K-1}\sum_{j = 2}^K (1-(1-p_r^j)^{N-1}(1-p_m^j)) \frac{(N-1)R_r+R_m}{N}\right) \\
    &\left(\sum_{k=\mathcal{M}_C}^{N-1} {N-1 \choose k} p_r^k(1-p_r)^{N-1-k} + p_m{N-1\choose \mathcal{M}_C-1} p_r^{\mathcal{M}_C-1}(1-p_r)^{N-\mathcal{M}_C} \right)
    \\
    & + \frac{(N-1)R_r+R_m}{K-1}\sum_{k=0}^{N-1} {N-1 \choose k} \frac{1}{k+1} \sum_{j = 2}^{K} p_m^j p_r^{jk}(1-p^r)^{N-1-k}\\
    &\left( 1- \left(\sum_{k=0}^{\mathcal{M}_C-2} {N-1 \choose k} p_r^k(1-p_r)^{N-1-k} + (1-p_m){N-1\choose \mathcal{M}_C-1} p_r^{\mathcal{M}_C-1}(1-p_r)^{N-\mathcal{M}_C} \right)^j\right).
\end{align*}

\section{Further illustration}

\begin{figure}
    \centering
    \includegraphics[width=\textwidth]{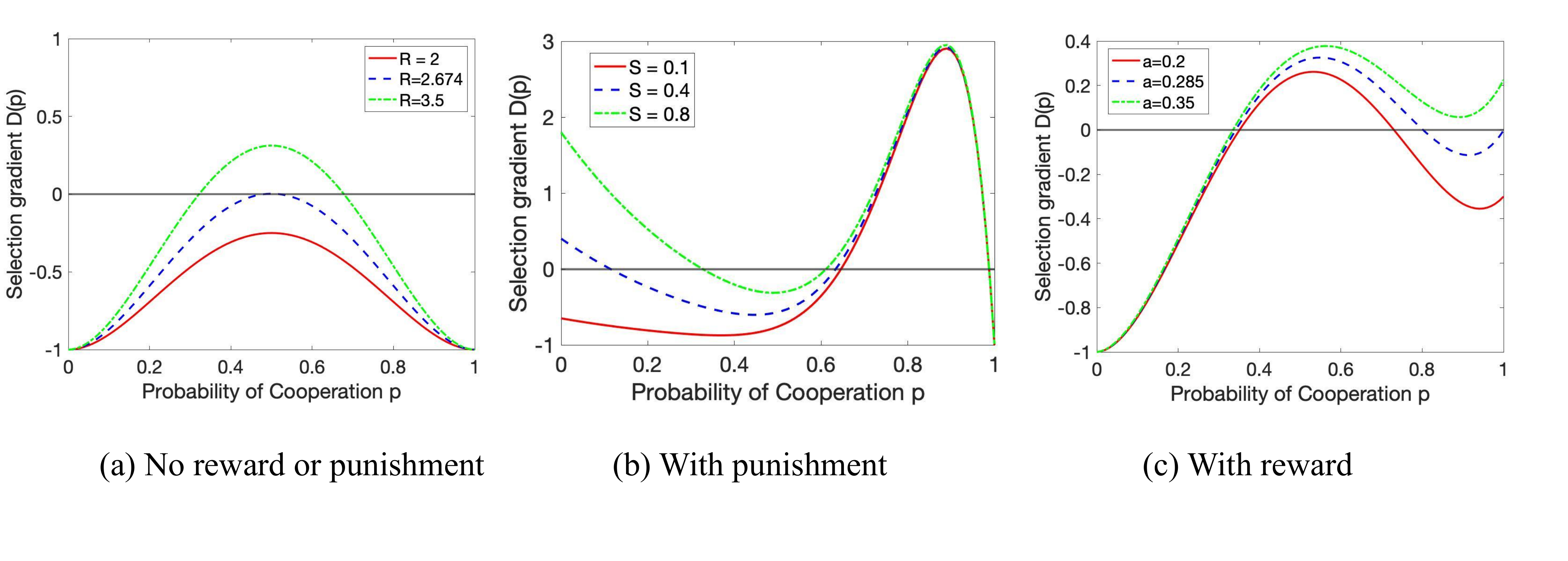}
    \caption{Selection gradient D(p) for (a) $N = 5, \mathcal{M}_C = 3, C = 1$, (b) $N = 10, \mathcal{M}_C = 8, C = 1, B = 10$, (c) $N = 5, \mathcal{M}_C = 3, C = 1, B = 3.5$}
    \label{fig:D_form}
\end{figure}

The selection gradient without incentives has a single maximum at $p=\frac{M_C-1}{N-1}$ and is negative for $p=0,1$. In Figure \ref{fig:D_form}, we see the emergence of a bifurcation. For $R<R^*\approx 2.674$, there exist no singular strategies, and the population always evolves to the antisocial equilibrium $p=0$. However, for $R>R*$, there exist two singular strategies $0<p_1^*<p_2^*<1$ and the population evolves to the social equilibrium $p_2^*$. Rewards and punishment can cause the emergence of additional equilibria. 

\begin{figure}
    \centering
    \includegraphics[width=.8\textwidth]{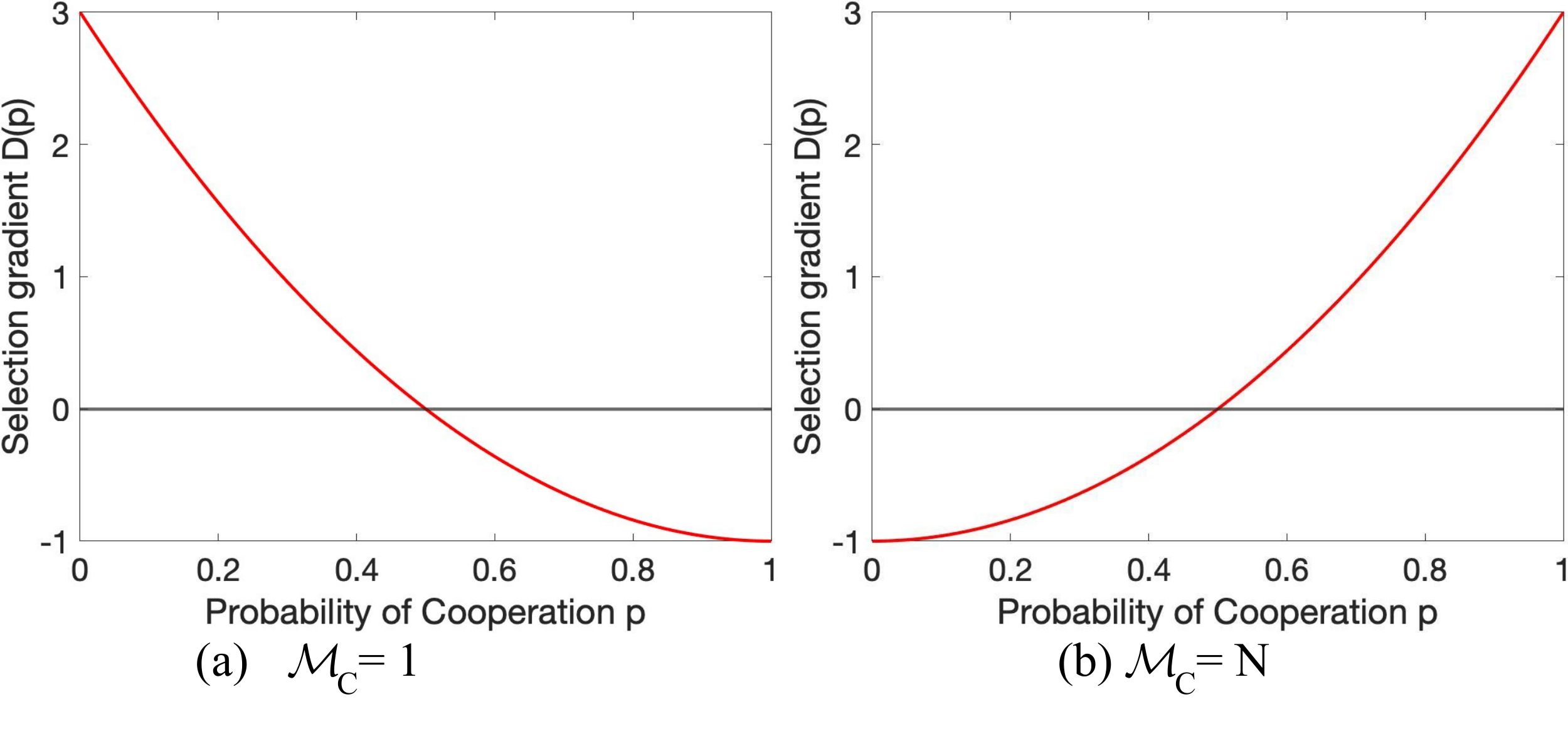}
    \caption{Equilibria in the public goods game $N = 3, \mathcal{M}_C = 1$ respectively $\mathcal{M}_C = 3, B = 4, C = 1$.}
    \label{fig:equi_special}
\end{figure}

For $\mathcal{M}_C = 1$, the maximum of the selection gradient $D$ is at $p=0$ (Figure \ref{fig:equi_special}(a)). Here, the selection gradient has a single root for any $C>0$. This root is an attractor and evolutionary branching can evolve in the model with incentives. However, for $\mathcal{M}_C=N$, the maximum is at $p=1$ (Figure \ref{fig:equi_special}(b)). The root of $D$ here is a repellor and we can never observe evolutionary branching.

\begin{figure}
    \centering
    \includegraphics[width=.98\textwidth]{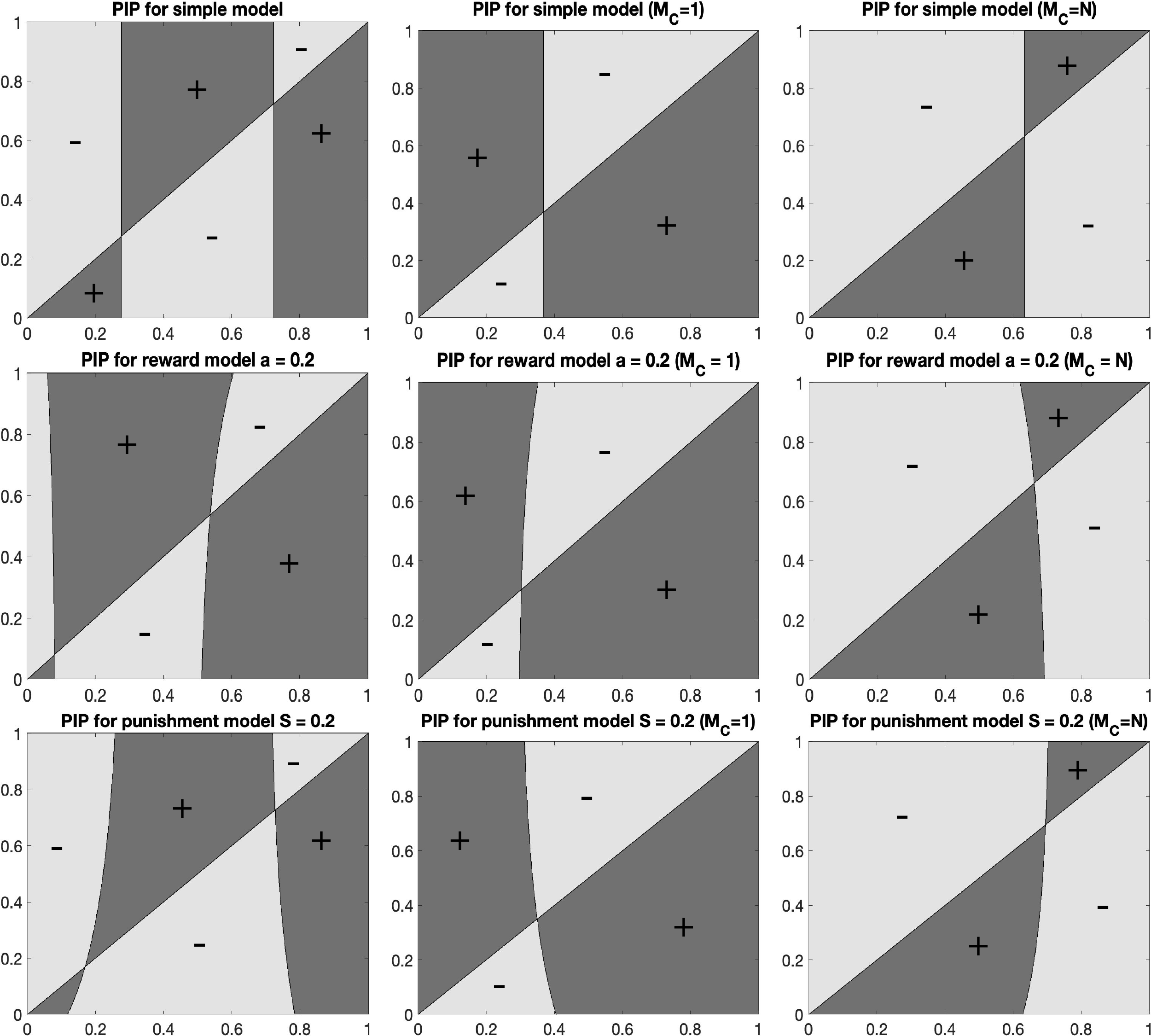}
    \caption{Pairwise invasibility plots for the model with rewarding and punishing. Parameters are $N = 3, B = 2.5, C = 1$ and $\mathcal{M}_C = 2$ respectively $\mathcal{M}_C = 1$ respectively $\mathcal{M}_C = 3$ for the simple model, and $N = 5, B = 4, C = 1$ and $\mathcal{M}_C = 3$ respectively $\mathcal{M}_C = 1$ respectively $\mathcal{M}_C = 5$ for the reward and punishment model.}
    \label{fig:pip}
\end{figure}

A further illustration of the analysis is given in Figure \ref{fig:pip}. Here, we see that in the simple model without incentives, every strategy has the same payoff in a population of $p_1^*$ or $p_2^*$ players, the invasion fitness is 0 on the vertical lines through the singular strategies. If rewards are introduced in the model, the lines curve to the right for punishment. So, for rewards, any other strategy can invade the population and we observe evolutionary branching. For punishment, on the other side, no other strategy has a higher payoff than the resident population for the convergent stable strategies.

\end{document}